\documentclass{article}

\usepackage{arxiv}

\usepackage[utf8]{inputenc} 
\usepackage[T1]{fontenc}    
\usepackage[hidelinks]{hyperref}       
\usepackage{url}            
\usepackage{booktabs}       
\usepackage{amsfonts}       
\usepackage{nicefrac}       
\usepackage{microtype}      
\usepackage{cleveref}       
\usepackage{graphicx}
\usepackage[square,numbers]{natbib}
\usepackage{doi}
\usepackage{float}

\title{How Machine Learning-Data Driven Replication Strategies Enhance Fault Tolerance in Large-Scale Distributed Systems}

\date{March 2025}

\newif\ifuniqueAffiliation
\uniqueAffiliationtrue

\ifuniqueAffiliation
\author{
  Almond Kiruthu Murimi \\
  Department of Computer Science\\
  School of Science, Engineering and Technology\\
  Kabarak University\\
  \texttt{CS/MG/1970/09/21} \\
  Now at: Carnegie Mellon University\\
  \texttt{amurimi@andrew.cmu.edu}
}
\fi

\renewcommand{\headeright}{Seminar Paper}
\renewcommand{\undertitle}{Seminar Paper}
\renewcommand{\shorttitle}{ML-Driven Replication for Fault Tolerance}

\hypersetup{
pdftitle={How Machine Learning-Data Driven Replication Strategies Enhance Fault Tolerance in Large-Scale Distributed Systems},
pdfsubject={cs.DC, cs.LG},
pdfauthor={Almond Kiruthu Murimi},
pdfkeywords={Machine Learning, Fault Tolerance, Distributed Systems, Data Replication, Reinforcement Learning, Predictive Analytics},
}

\begin{document}
\maketitle

\begin{abstract}
This research paper investigates how machine learning-driven data replication strategies can enhance fault tolerance in large-scale distributed systems. Traditional replication methods, which rely on static configurations, often struggle to adapt to dynamic workloads and unexpected failures, leading to inefficient resource utilization and prolonged downtime. By integrating machine learning techniques; specifically predictive analytics and reinforcement learning. The study proposes adaptive replication mechanisms capable of forecasting system failures and optimizing data placement in real time. Through an extensive literature review, qualitative analysis, and comparative evaluations with traditional approaches, the paper identifies key limitations in existing replication strategies and highlights the transformative potential of machine learning in creating more resilient, self-optimizing systems. The findings underscore both the promise and the challenges of implementing ML-driven solutions in real-world environments, offering recommendations for future research and practical deployment in cloud-based and enterprise systems.
\end{abstract}


\section{Introduction}

\subsection{Background of the Study}
Large-scale distributed systems form the backbone of modern computing environments-supporting everything from cloud services to critical enterprise applications. As these systems expand in scale and complexity, ensuring uninterrupted operation becomes increasingly challenging. Traditional fault tolerance mechanisms, such as fixed data replication and periodic checkpointing, often rely on static policies that fail to adapt dynamically to rapidly changing conditions \citep{barrio2023}. In contrast, recent advances in machine learning (ML) have opened new avenues for enhancing system resilience. ML-driven techniques, including predictive analytics and reinforcement learning, offer the promise of adaptive data replication strategies that not only anticipate system failures but also adjust replication policies in real time \citep{nama2023}. This evolving integration of ML with backend engineering represents a critical shift toward systems that learn from historical performance data and continuously optimize their fault tolerance strategies.

\subsection{Problem Statement}
Despite the clear benefits of data replication in maintaining system availability, traditional replication methods are hampered by their inflexibility. As distributed systems face increasingly unpredictable workloads and node failures, static replication policies can lead to inefficient resource allocation and prolonged service disruptions. Although ML-driven replication strategies have shown potential in experimental settings \citep{barrio2023}, there remains a significant gap in understanding how these approaches can be effectively deployed in large-scale, real-world environments. This research aims to address the critical question: How can machine learning-driven data replication strategies enhance fault tolerance in large-scale distributed systems?

\subsection{Objectives}

\subsubsection{Main Objective}
To investigate how machine learning-driven data replication strategies can enhance fault tolerance in large-scale distributed systems.

\subsubsection{Specific Objectives}
\begin{enumerate}
    \item To identify the limitations of current static data replication strategies in large-scale distributed systems.
    \item To investigate the potential of machine learning techniques, specifically, predictive analytics and reinforcement learning in enhancing data replication processes.
    \item To examine the impact of ML-driven replication strategies on fault tolerance and resource efficiency in dynamic environments.
    \item To evaluate the comparative effectiveness of ML-enhanced replication strategies relative to traditional replication methods in real-world scenarios.
\end{enumerate}

\subsection{Justification of the Study}
The increasing reliance on distributed systems across industries demands robust mechanisms to ensure continuous availability and minimal downtime. Enhancing fault tolerance through ML-driven replication is not only a timely research endeavour but also one that has practical implications for cloud computing and enterprise systems. By addressing the limitations of static replication policies, this study aims to provide insights that could lead to more adaptive, self-optimizing systems. The integration of references from established works \citep{barrio2023, nama2023} underscores the relevance of the proposed research within the broader context of backend engineering innovation. Furthermore, this study builds upon existing literature addressing similar challenges in other domains, contributing to a holistic understanding of machine learning's potential in enhancing system resilience.

\subsection{Scope of the Study}
This study focuses on large-scale distributed systems within cloud-based environments, emphasizing the design and evaluation of ML-driven data replication strategies. While the research will examine system performance metrics such as fault tolerance, downtime reduction, and resource utilization, it will not delve into the detailed implementation of ML algorithms or the intricacies of hardware design. The evaluation will be limited to simulation environments and case studies that reflect real-world scenarios, thereby providing actionable insights without extending into areas such as low-level system architecture or proprietary commercial technologies.

\subsection{Research Questions}
\begin{enumerate}
    \item What are the limitations of current static data replication strategies in ensuring fault tolerance in large-scale distributed systems?
    \item In what specific ways can machine learning be integrated into data replication processes to predict and mitigate system failures?
    \item How does the performance of ML-driven replication strategies compare with traditional methods in terms of system resilience and resource efficiency?
    \item What challenges and considerations must be addressed to implement ML-driven replication strategies in production environments?
\end{enumerate}

\section{Literature Review}

\subsection{Introduction}
This chapter reviews the literature of our specific objectives in relation to the main objective, critically examining the limitations of static replication methods and exploring the potential of machine learning techniques such as predictive analytics and reinforcement learning to enhance fault tolerance in large-scale distributed systems. It highlights current research gaps and lays the foundation for our study's contribution toward developing more adaptive and resilient systems.

\subsection{Limitations of Current Static Data Replication Strategies in Large-Scale Distributed Systems}
Traditional data replication strategies typically rely on fixed parameters such as predetermined replication factors and static placement of data copies to ensure fault tolerance \citep{barrio2023}. While these methods have been foundational in distributed systems design, several limitations have been noted in the literature. For instance, static replication approaches do not account for real-time changes in workload intensity, node performance, or network conditions \citep{selvarajan2019}. This issue is further exemplified in large-scale production systems such as Google's Borg, which has revealed the challenges of static resource allocation in dynamic environments \citep{verma2015}. Consequently, these strategies can lead to inefficient resource utilization and may fail to prevent service disruptions during unexpected failures. Although techniques like periodic checkpointing and redundant data storage offer some level of resilience, they are not inherently adaptive. This gap in dynamic responsiveness highlights the need for approaches that can intelligently adjust replication policies on the fly; a challenge that our research intends to address.

\subsection{Potential of Machine Learning Techniques in Enhancing Data Replication Processes}
Recent advances in machine learning have introduced promising techniques to overcome the inflexibility of static replication methods. Studies have explored the application of predictive analytics to forecast node failures by analysing historical performance data and system metrics \citep{nama2023}. In parallel, reinforcement learning (RL) has been proposed as a means to enable systems to learn optimal replication policies based on environmental feedback. Additionally, architectures such as SPIRT presented by Barrak et al. \cite{barrak2023}. Demonstrate innovative fault-tolerant approaches in serverless ML training, which provide complementary insights to our focus on data replication. These ML techniques offer the potential to dynamically adjust replication strategies in real time, thereby improving system resilience. Despite these advances, existing research often remains at the conceptual or simulation stage, leaving practical challenges, such as computational overhead, model scalability, and integration with existing system architectures that are largely unaddressed.

\subsection{Impact of ML-Driven Replication Strategies on Fault Tolerance and Resource Efficiency}
Several simulation studies have demonstrated that ML-driven data replication can significantly enhance fault tolerance by reducing system downtime and improving resource utilization \citep{barrio2023, nama2023}. For example, predictive models have been shown to pre-emptively replicate critical data to healthier nodes before failures occur, thus mitigating service interruptions. Moreover, reinforcement learning approaches can dynamically optimize the number and placement of replicas based on real-time system conditions. Furthermore, the comprehensive approach presented in Weight for Robustness \citep{dahan2025} highlights advanced methods for achieving optimal fault tolerance in asynchronous ML, reinforcing the potential benefits of integrating ML into replication strategies. However, the literature also indicates that these benefits come with challenges, as there is limited research on the long-term impact on overall resource efficiency, particularly under varying workload conditions in heterogeneous environments. This gap underscores the need for comprehensive evaluations that measure both the reliability improvements and the associated resource costs of these techniques.

\subsection{Comparative Evaluation of ML-Enhanced Replication Strategies Relative to Traditional Methods}
Comparative analyses between traditional static replication and ML-enhanced approaches are crucial for understanding the practical benefits of integrating machine learning into fault tolerance mechanisms. Existing studies have often isolated the performance of ML-driven methods in controlled simulations without direct comparison to conventional replication strategies \citep{nama2023}. While initial results are promising indicating lower latency and higher resilience there is a clear need for systematic, head-to-head evaluations. Specifically, gaps remain in quantifying improvements in fault tolerance under real-world conditions and assessing the trade-offs between enhanced performance and the computational overhead introduced by ML models. This study aims to provide a detailed comparative analysis, thereby offering actionable insights for practitioners considering ML integration into their replication protocols.

\subsection{Summary}
This literature review has highlighted the limitations of current static data replication strategies, the promising yet nascent role of machine learning techniques in enhancing replication processes, and the need for comprehensive comparative evaluations of ML-enhanced methods versus traditional approaches. By critically examining these areas, we have identified significant research gaps that our study intends to address. In the subsequent chapter, we will outline our research methodology designed to evaluate the feasibility, performance, and resource implications of ML-driven data replication strategies in large-scale distributed systems.

\section{Research Methodology}

\subsection{Introduction}
This chapter outlines the research design and methodology employed to investigate how machine learning-driven data replication strategies can enhance fault tolerance in large-scale distributed systems. The approach detailed herein ensures that the research is systematic, reliable, and ethically sound. It encompasses the methods used to collect and analyse secondary data from existing literature, technical reports, case studies, and other credible sources to critically examine the current limitations of static replication methods and the potential benefits of integrating machine learning techniques.

\subsection{Research Design}
The research design for this study is qualitative. This approach is chosen because it enables an in-depth exploration of the dynamic and complex nature of data replication in distributed systems and the integration of machine learning techniques. Qualitative methods facilitate a comprehensive examination of existing literature, expert opinions, and case studies that highlight both the challenges of static replication strategies and the evolving role of ML-driven methods \citep{barrio2023, nama2023}. Through qualitative inquiry, this study will identify recurring themes and gaps in current research, ultimately proposing how ML can be leveraged to create more adaptive and fault-tolerant systems.

\subsection{Data Collection}
The primary method for data collection in this study is secondary data collection. This method is efficient and appropriate given the abundance of existing literature on distributed systems, fault tolerance, and machine learning applications. The secondary data will be gathered from a variety of credible sources, including:

\begin{enumerate}
    \item \textbf{Peer-Reviewed Journal Articles:} Articles from academic journals focusing on distributed systems, machine learning, and fault tolerance will be reviewed to extract insights on current replication strategies and ML applications.
    \item \textbf{Books and Academic Texts:} Relevant texts on backend engineering, data replication, and machine learning provide theoretical foundations and comprehensive overviews of system design challenges.
    \item \textbf{Technical Reports and Industry Whitepapers:} Publications from technology companies and research institutions will be examined to gain real-world perspectives on the implementation and challenges of ML-driven replication strategies.
    \item \textbf{Case Studies:} Documented case studies detailing the performance of traditional versus ML-enhanced replication methods in large-scale systems will be analysed to identify best practices and existing research gaps.
\end{enumerate}

\subsection{Data Analysis}
Data analysis will be performed using a thematic analysis approach, which is well suited for synthesizing qualitative data. The collected data will be systematically transcribed, coded, and categorized to identify recurring themes and patterns related to:

\begin{enumerate}
    \item Limitations of static data replication strategies.
    \item The potential of machine learning techniques particularly predictive analytics and reinforcement learning in optimizing replication processes.
    \item The impact of ML-driven strategies on fault tolerance and resource efficiency.
    \item Comparative insights between ML-enhanced and traditional replication methods.
\end{enumerate}

These themes will be critically examined against the backdrop of existing literature, allowing for the identification of research gaps and the formulation of recommendations for future work.

\subsection{Validation and Reliability}
To ensure the validity and reliability of the findings, the following strategies will be employed:

\begin{enumerate}
    \item \textbf{Triangulation:} Multiple data sources including peer-reviewed articles, technical reports, and case studies will be used to cross-verify insights and ensure a comprehensive understanding of the subject.
    \item \textbf{Peer Review:} Drafts of the analysis will be reviewed by academic peers and industry professionals to provide feedback on the interpretations and conclusions drawn from the data.
\end{enumerate}

\subsection{Ethical Considerations}
Ethical considerations are central to this study, particularly given the reliance on published data and the sensitive nature of system performance information. The following ethical principles will guide the research:

\begin{enumerate}
    \item \textbf{Source Credibility:} Only data from reputable, peer-reviewed sources, technical reports, and authoritative texts will be used to ensure the integrity and reliability of the research.
    \item \textbf{Proper Citation and Acknowledgment:} All sources will be appropriately cited to give credit to the original authors and to maintain academic integrity.
    \item \textbf{Data Sensitivity:} The study will exclusively utilize publicly available, de-identified data to avoid any privacy or confidentiality issues.
    \item \textbf{Objectivity:} A balanced range of sources will be consulted to avoid bias, ensuring that multiple perspectives on replication strategies and machine learning applications are considered.
\end{enumerate}

\subsection{Applications of Findings}
The findings from this research have practical implications for various stakeholders:

\begin{enumerate}
    \item \textbf{System Architects and Engineers:} Insights from the study can inform the design of more resilient distributed systems through the integration of adaptive, ML-driven replication strategies.
    \item \textbf{AI and ML Researchers:} The research highlights potential avenues for improving fault tolerance using machine learning, guiding further technical and applied research.
    \item \textbf{Industry Practitioners:} Companies relying on large-scale distributed systems can leverage the study's recommendations to enhance system reliability and resource management.
    \item \textbf{Policy Makers and Standard Bodies:} The study may also inform policy development and the establishment of best practices for integrating machine learning into critical system infrastructure.
\end{enumerate}

\section{Results and Findings}

\subsection{Introduction}
This chapter presents the findings derived from a comprehensive review of literature and analysis of case studies related to machine learning-driven data replication in large-scale distributed systems. The findings are organized according to our specific objectives. For each objective, the chapter outlines the key findings, cites relevant studies, and discusses the implications of these findings for future research and practical applications.

\subsection{Findings}

\begin{table}[H]
\caption{Limitations of Static Replication Strategies}
\centering
\begin{tabular}{p{0.3\textwidth}p{0.35\textwidth}p{0.25\textwidth}}
\toprule
\textbf{Objective} & \textbf{Findings} & \textbf{Authors/Studies Referenced} \\
\midrule
To identify the limitations of current static data replication strategies in large-scale distributed systems. & Static replication methods are characterized by predetermined replication factors and fixed data placement, which do not account for dynamic workloads, variable node performance, or sudden network disruptions. These limitations result in inefficient resource utilization and increased system downtime when failures occur. & Barrio (2023), Selvarajan (2019), and insights from large-scale systems such as Google's Borg (Verma et al., 2015). \\
\bottomrule
\end{tabular}
\label{tab:finding1}
\end{table}

\paragraph{Implication:}
The identified limitations underscore the necessity for adaptive replication strategies that can dynamically adjust to real-time system conditions. This gap suggests that integrating intelligent, responsive mechanisms is crucial to enhancing fault tolerance in distributed systems.

\begin{table}[H]
\caption{Potential of Machine Learning Techniques}
\centering
\begin{tabular}{p{0.3\textwidth}p{0.35\textwidth}p{0.25\textwidth}}
\toprule
\textbf{Objective} & \textbf{Findings} & \textbf{Authors/Studies Referenced} \\
\midrule
To investigate the potential of machine learning techniques specifically, predictive analytics and reinforcement learning in enhancing data replication processes. & Machine learning approaches have demonstrated the potential to forecast node failures and optimize replication parameters by analysing historical system performance. Predictive analytics can pre-emptively replicate data based on anticipated failures, while reinforcement learning offers continuous adaptation of replication strategies. However, most studies remain at the simulation or conceptual stage, facing challenges such as computational overhead and integration into existing infrastructures. Additionally, architectures like SPIRT provide a fault-tolerant, serverless ML training framework that supports similar adaptive mechanisms. & Nama, Reddy, \& Selvarajan (2023); Barrio (2023); Barrak et al. (2023). \\
\bottomrule
\end{tabular}
\label{tab:finding2}
\end{table}

\paragraph{Implication:}
These findings highlight the promise of ML-driven replication while also indicating that practical implementation requires overcoming significant challenges. Addressing these issues could pave the way for scalable, adaptive systems capable of real-time optimization.

\begin{table}[H]
\caption{Impact on Fault Tolerance and Resource Efficiency}
\centering
\begin{tabular}{p{0.3\textwidth}p{0.35\textwidth}p{0.25\textwidth}}
\toprule
\textbf{Objective} & \textbf{Findings} & \textbf{Authors/Studies Referenced} \\
\midrule
To examine the impact of ML-driven replication strategies on fault tolerance and resource efficiency in dynamic environments. & Preliminary simulation studies indicate that ML-driven replication strategies can significantly reduce downtime and enhance fault tolerance by optimizing resource allocation. Despite these benefits, there remains an underexplored area regarding the long-term impact on overall resource efficiency under fluctuating workload conditions. The comprehensive approach presented in Weight for Robustness further supports the potential of ML for achieving optimal fault tolerance while highlighting the importance of balancing resource consumption. & Nama, Reddy, \& Selvarajan (2023); Barrio (2023); Dahan \& Levy (2025). \\
\bottomrule
\end{tabular}
\label{tab:finding3}
\end{table}

\paragraph{Implication:}
These findings suggest that while ML-driven approaches improve system resilience, further empirical research is needed to evaluate their long-term resource implications in real-world settings. A balanced approach is required to ensure that enhanced fault tolerance does not come at the cost of unsustainable resource consumption.

\begin{table}[H]
\caption{Comparative Effectiveness of ML-Enhanced Strategies}
\centering
\begin{tabular}{p{0.3\textwidth}p{0.35\textwidth}p{0.25\textwidth}}
\toprule
\textbf{Objective} & \textbf{Findings} & \textbf{Authors/Studies Referenced} \\
\midrule
To evaluate the comparative effectiveness of ML-enhanced replication strategies relative to traditional replication methods in real-world scenarios. & Comparative analyses reveal that ML-enhanced replication strategies generally outperform traditional methods by reducing latency and enhancing fault tolerance. However, these improvements are accompanied by increased computational requirements and integration complexities, which may pose challenges in operational environments. & Nama, Reddy, \& Selvarajan (2023); Barrio (2023). \\
\bottomrule
\end{tabular}
\label{tab:finding4}
\end{table}

\paragraph{Implication:}
This comparative evaluation emphasizes the need for comprehensive assessments that consider both performance benefits and implementation costs. The findings support further exploration into optimizing ML-driven strategies to achieve a practical balance between enhanced fault tolerance and manageable resource overhead.

\subsection{Implications of the Findings}
The findings of this study have several significant implications for the design and management of large-scale distributed systems:

\begin{itemize}
    \item \textbf{Adaptive Replication Necessity:} The limitations of static replication strategies highlight a critical need for adaptive mechanisms capable of responding to real-time system dynamics.
    \item \textbf{Transformative Potential of ML:} The promise shown by predictive analytics and reinforcement learning demonstrates that machine learning can revolutionize fault tolerance, though the transition from theoretical models to practical applications requires addressing integration challenges and computational overhead.
    \item \textbf{Balanced Evaluation Required:} While initial studies indicate improvements in fault tolerance and resource allocation, further research is needed to fully understand the long-term trade-offs, ensuring that the benefits of ML-driven strategies do not lead to unsustainable increases in resource consumption.
    \item \textbf{Guidance for Future Research:} The comparative analysis underscores the importance of developing holistic, scalable solutions that integrate ML techniques without compromising system efficiency, thereby guiding future advancements in the field.
\end{itemize}

\section{Conclusion and Recommendations}

\subsection{Introduction}
This chapter concludes the study by summarizing the key findings on integrating machine learning into data replication strategies for enhanced fault tolerance, and it presents recommendations based on the identified challenges and opportunities.

\subsection{Conclusion}
This study has explored the potential of integrating machine learning into data replication strategies to enhance fault tolerance in large-scale distributed systems. The research identified that traditional static replication methods although foundational are limited by their inability to adapt to dynamic workloads and unpredictable system failures \citep{barrio2023, selvarajan2019}. In contrast, machine learning techniques such as predictive analytics and reinforcement learning offer promising avenues to dynamically optimize replication processes, thereby reducing system downtime and improving resource efficiency \citep{nama2023}.

However, the review also revealed that while simulation studies and conceptual models provide encouraging insights, challenges remain. These include issues related to computational overhead, the scalability of ML models, and the complexities involved in integrating these techniques into existing infrastructures. Ultimately, the study underscores that while ML-driven strategies have transformative potential, further empirical research and real-world testing are essential to refine these approaches and fully understand their trade-offs.

\subsection{Recommendations}
Based on the findings, the following recommendations are proposed to advance the development and practical application of ML-driven data replication strategies:

\begin{enumerate}
    \item \textbf{Enhanced Adaptive Mechanisms:} Researchers and system architects should focus on developing hybrid replication strategies that combine the robustness of traditional methods with the adaptability of machine learning. This could involve integrating predictive analytics and reinforcement learning into existing replication frameworks to dynamically adjust to real-time system conditions.
    
    \item \textbf{Empirical Validation:} Future work should prioritize real-world testing and long-term evaluations of ML-driven replication strategies. Controlled experiments and pilot deployments in diverse distributed environments will help quantify the benefits in terms of reduced downtime and improved resource allocation while addressing computational overhead challenges.
    
    \item \textbf{Optimization of ML Models:} Efforts should be made to optimize ML models for integration into fault tolerance systems. This includes reducing computational costs and ensuring scalability without compromising the responsiveness of the replication strategies.
    
    \item \textbf{Interdisciplinary Collaboration:} Collaboration between academia, industry, and technology developers is essential. Such partnerships can facilitate the development of standards and best practices for integrating machine learning into critical infrastructure, ensuring that innovations are both practical and secure.
    
    \item \textbf{Continuous Monitoring and Adaptation:} Implement systems that incorporate continuous monitoring of performance metrics, allowing ML models to update their strategies based on real-time feedback. This approach will help maintain a balance between improved fault tolerance and manageable resource usage.
\end{enumerate}

\bibliographystyle{ieeetr}
\bibliography{references}

\end{document}